# Holographic Astigmatic Particle Tracking Velocimetry (HAPTV)

Zhou Zhou[1], Santosh Kumar Sankar[2], Kevin Mallery[2], Wensheng Jiang[1,3], Jiarong Hong[2]*

[1] Physical Oceanography Laboratory/CIMST, Ocean University of China and Qingdao National Laboratory for Marine Science and Technology, Qingdao 266100, China
[2] Department of Mechanical Engineering, University of Minnesota, 111 Church St SE, Minneapolis, MN 55455, USA
[3] Laboratory of Marine Environment and Ecology, Ocean University of China, Qingdao 266100, China
*corresponding author, Email: jhong@umn.edu

## Abstract

The formation of twin images in digital inline holography (DIH) prevents the placement of the focal plane in the center of a sample volume for DIH-based particle tracking velocimetry (DIH-PTV) with a single camera. As a result, it is challenging to apply DIH-PTV for flow measurements in large-scale laboratory facilities or many field applications where it would otherwise be desirable due to the low cost and compact setup. Here we introduce holographic astigmatic PTV (HAPTV) by inserting a cylindrical lens in the optical setup of DIH-PTV, generating distorted holograms. Such distortion is subsequently utilized in a customized reconstruction algorithm to distinguish tracers positioned on different sides of the focal plane which can in turn be placed in the middle of a sample volume. Our HAPTV approach is calibrated under high (1 μm/pixel) and low (10 μm/pixel) magnifications with an error standard deviation of 4.2 μm (one particle diameter) and 120.7 μm (~5 times the particle diameter), respectively. We compare the velocity field of a laminar jet flow obtained using HAPTV and conventional PIV to illustrate the accuracy of the technique when applied to practical flow measurement applications. The work demonstrates that HAPTV improves upon the depth of field of conventional astigmatic PTV and enables the implementation of DIH-based PTV for in situ applications.

## 1. Introduction

Velocity measurements in a field environment are usually conducted using single-point probing techniques such as sonic anemometry [1] in air and acoustic Doppler velocimetry (ADV) [2] in water due to their simplicity and robustness. However, these techniques can only provide limited information of the velocity field (i.e., single-point measurements), and require the presence of sufficient scatterers (e.g., 3.4-5.0×10$^4$ particles/cm$^3$ for 5 MHz ADV) [3] in the fluid medium, preventing their applications in pristine environments such as the deep ocean [4,5]. In comparison, three-dimensional (3D) flow field imaging techniques such as 3D particle image/tracking velocimetry (PIV/PTV) [6,7], commonly used in the lab settings, can yield unprecedented information for fundamental understanding of many important physical processes in the atmosphere and ocean. Although multi-camera 3D PIV/PTV techniques such as tomographic PIV [8–10] and synthetic aperture PTV [11] are more common, several single-camera 3D PTV techniques using plenoptic [12,13], defocusing [14–16], and holographic [17–20] imaging approaches have been gaining popularity for velocity measurement due to their simple setup and minimal calibration requirements [14,21–24]. Such advantages allow these techniques to be applied underwater and reduce the intrusiveness to the measurement volume, which is crucial for in situ applications.

In defocusing PTV, a multi-hole aperture is commonly used for generating the defocused image [25–27] where the shape variation of then object (i.e., degree of defocusing) as it moves away from the focal plane encodes the depth position , without any other specialized optical components. Objects that are not located on the focal plane produce multiple images on the sensor, one for each pinhole, and the distance between these images encodes the depth of the object. However, due to its multi-image encoding of each particle, the method is limited to low tracer concentration (e.g., 0.12 particles/cm$^3$) [14] as particle localization becomes challenging as the number of tracers in the sample volume increases. Furthermore, the use of a pinhole aperture significantly reduces the amount of light collected by the sensor [26], demanding the usage of high-power illumination which increases the cost and lowers the reliability of such systems for in situ measurements [28].

Another way to implement defocusing PTV which overcomes the limitation of the traditional defocusing method employs astigmatism, i.e. non-axisymmetric characteristics of the optical system which cause light rays propagating on different planes to focus at different axial locations. A higher degree of astigmatism separates the focal planes by a greater distance, thereby introducing additional distortion. In astigmatic PTV (APTV), astigmatism can be introduced by mounting a cylindrical lens between the imaging lens and the camera sensor [29,30], tilting the camera by a small angle to the measurement plane [31], or using a spherical wave illumination source offset from the optical axis [32]. Such astigmatic images of spherical particles appear elliptical, with their degree of ellipticity dependent on the distance of the object from each focal plane [33]. By measuring the major and minor axes of the ellipse, the depth of the particle can be uniquely determined [29,30]. Apart from high signal-noise ratio relative to conventional defocusing PTV, APTV can handle higher tracer concentrations (e.g., 640 particles/cm$^3$) [34] as the signal from each object is confined to a single position on the image. However, APTV suffers from several limitations that prevent its adoption for in situ measurements. First, the technique has only been implemented for spherical particles [28] and extending it to non-spherical particles found in natural environments will be challenging. In addition, a limited depth of field (~2 mm) [35] and low depth sensitivity have limited the technique primarily to microscopic applications.

A final single-camera approach is holographic PTV which uses the interference between the light scattered from objects and a reference light to encode their depth information into holograms. Such encoding is convolutional and compressive, and the depth information can be extracted through a reconstruction process using a diffraction formulation. Therefore, holographic PTV has a higher depth sensitivity and can deal with higher tracer concentrations compared with defocusing approaches because the compressive nature of holography means that the image contains denser information. In particular, holographic PTV using an inline setup (DIH-PTV) – which includes laser, camera, and lens aligned along a single optical axis – requires very low laser power and is highly compact, suitable for in situ measurements [19,22]. For typical in situ applications, we must place the laser and imaging systems sufficiently far apart to avoid introducing external disturbances (from the enclosure) to the flow measurements. Under such a measurement scenario, it becomes challenging to capture particles located far from the image plane, due to the drop in the signal strength (i.e., the scattered intensity) with distance from the sensor. Such a limitation can be overcome by simply positioning the imaging plane inside the measurement volume. However, the numerical reconstruction of in-line holograms which are usually based on the recorded intensity distributions do not include any phase information that can help differentiate objects on either side of the hologram. This is known as the twin image problem. To tackle the twin image problem, several methods have been proposed for distinguishing the objects on either side of the imaging plane in DIH. For example, the phase shifting method [36,37] varies the phase of the reference beam to reconstruct the complex amplitude of an object. The recorded objects must remain stationary while the reference phase is varied. Thus, the phase shifting method is only applicable for static or slowly-moving objects. Phase shifting is also mechanically complex, making this approach unsuitable for in situ measurements. Ling et al. [38] uses a DIH setup with two synchronized cameras to record holograms at two planes separated by a short distance and determine the side of an object with respect to the focal plane by comparing the object images in the two holograms. However, such a method increases the complexity of the DIH system in terms of the setup and calibration, limiting its applications for in situ measurements.

In this study, we present a new method, termed holographic astigmatic particle tracking velocimetry (HAPTV), which combines the advantages of both APTV and DIH-PTV to solve the twin image problem and acquire 3D flow fields in a sample volume with an extended depth. Section 2 provides a description of the general methodology of HAPTV, followed by the calibration of our method presented in Section 3. In Section 4, we apply the calibrated HAPTV to the measurement of a jet flow and compare to a PIV measurement. The conclusion of the work is provided in Section 5.

## 2. Methodology

### 2.1. General principle of HAPTV

Figure 1 presents a schematic of our system where we introduce a planar cylindrical lens between the camera and imaging lens of a conventional digital in-line holographic setup, as a source of astigmatism. We orient the axis of curvature of the cylindrical lens along the horizontal direction, resulting in the focal length along the $y$ axes being altered while the one along $x$ remains the same (i.e. as it is without the cylindrical lens). This configuration leads to particle fringe patterns appearing elongated when out of focus. Such an elongation encodes the direction of the particle from the image plane, with the orientation changing from horizontal when the particle is on the laser side of the focal plane ($z<0$) to vertical when the particle is on the camera side ($z>0$) as illustrated in figure 1. This direction encoding solves the twin image problem associated with DIH.

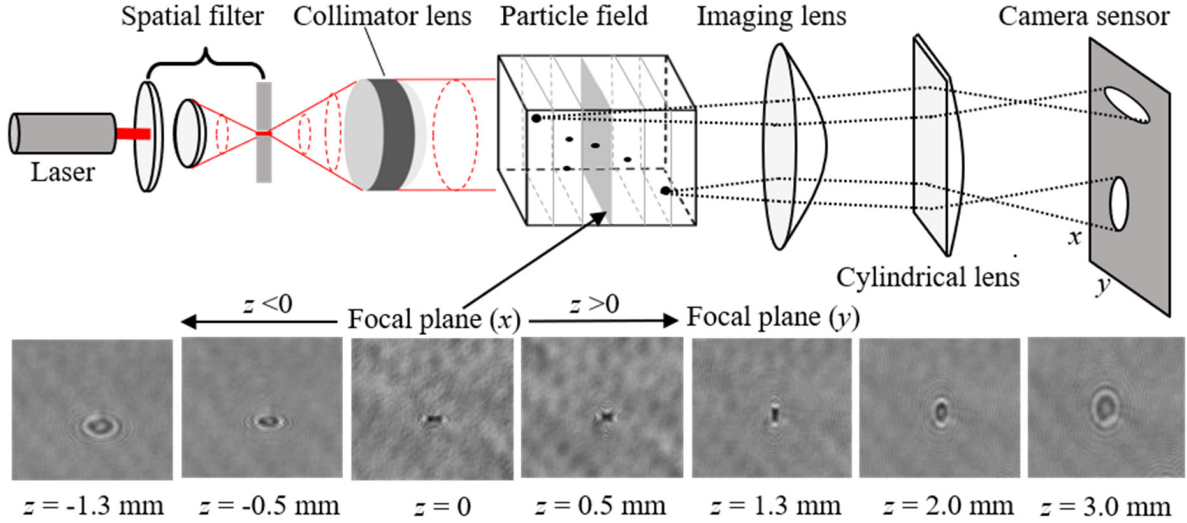

**Figure 1**. Schematic of the HAPTV optical setup (upper panel) and images of a 25 μm polystyrene particle at various $z$ distances from the focal plane (lower panel).

## 2.2. Digital reconstruction and 3D localization of objects

In this study, all recorded astigmatic holograms are processed using minor variations on standard holographic processing routines in order to extract the position information, including steps of image enhancement, reconstruction, and centroid calculation. The first step of the algorithm involves enhancement of the hologram through an ensemble-averaged background subtraction in order to eliminate stationary artifacts as well as increase the signal-to-noise ratio (SNR) of the fringe patterns. Next, the enhanced holograms (figure 2a) are numerically reconstructed through convolution with a Rayleigh-Sommerfeld diffraction kernel (also known as the angular spectrum method [39]) resulting in a 3D intensity field as shown in figure 2(b). The volume illustrates the unique signature of a reconstructed astigmatic hologram of a spherical particle, where the object is refocused along the horizontal and vertical directions at two separate planes, with the relative order of the planes dependent on the initial direction of elongation, as seen in figure 1. To better capture this signature, we divide the centroid calculation step into an in-plane localization and a depth localization substeps.

**In-plane localization:** A maximum intensity projection of the 3D intensity field along the longitudinal ($z$) direction is first calculated to generate a 2D image with all particles appearing as bright crosses with their $xy$ position at the cross centers (figure 2c). Next, we employ cross correlation with a cross-shaped template (figure 2d), the shape for which has been optimized for efficient extraction and minimizing ghost particles, to identify the center of each particle. The peak of the cross-correlation map (figure 2e) is extracted as an intensity weighted centroid through a manual thresholding operation as shown in figure 2(f) and used as the $xy$ position of the particle.

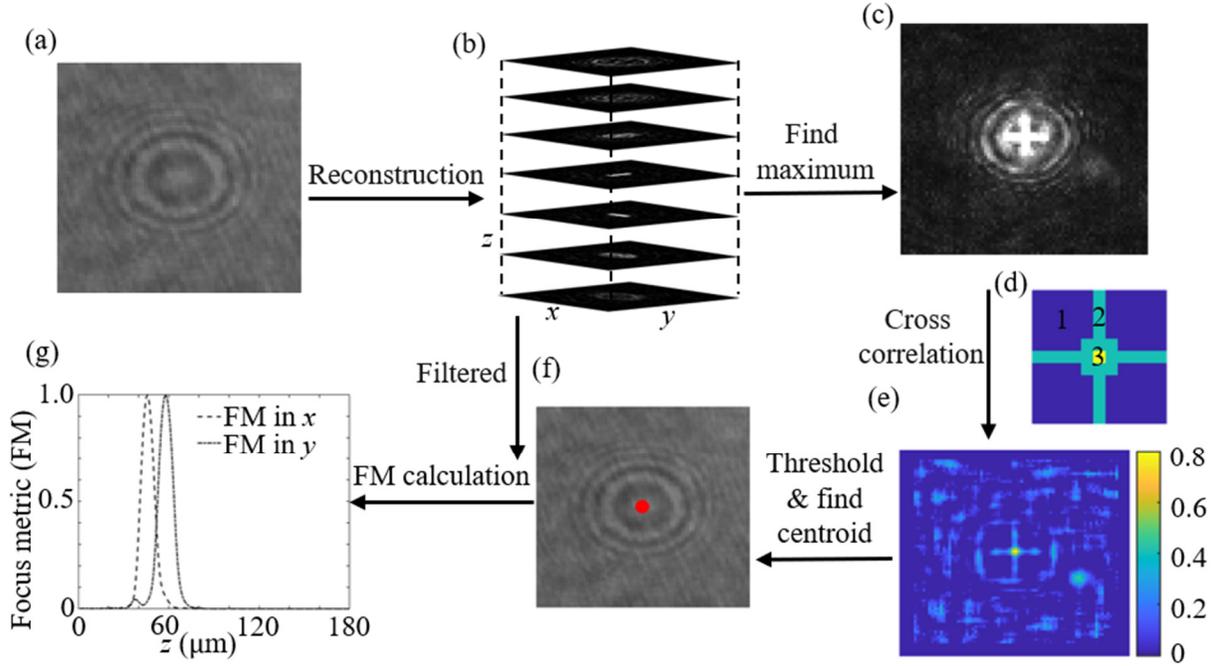

**Figure 2.** The flow chart of the processing procedure to extract 3D positions of particles recorded using holographic astigmatic imaging setup.

**Depth localization:** Once the particle position in *xy* is identified, we then determine the depth using an intensity-based focus metric. The sharp edges of the two focal planes are accentuated by applying a Sobel filter to the reconstructed intensity field. The operation is performed separately in both the *x* and *y* directions, producing two filtered volumes. Next, we calculate longitudinal (along *z* axes) profiles of the squared sum of the filtered intensity within two bounding boxes along each of the axes. Each bounding box has a ~4:1 aspect ratio, matching the elongated shape of the particle when in focus. We also include an inverse distance weighting of the intensity along the shorter axes, centered at the middle of the window to further enhance the signal intensity. The peaks of the longitudinal profiles are identified with sub-pixel accuracy using a spline interpolation. We define the depth of the particle using the position of the *x* peak (i.e. when the object is most focused as horizontal line), since the focal plane along this direction is insensitive to the presence of the cylindrical lens as demonstrated in Section 2.1. Finally, the direction of the particle (positive or negative) is identified by the relative order of the peaks. If the particle focuses along x first then the particle is located on the laser side of the image plane (*z*<0). This case is shown in figure 2(g). Conversely, if the particle focuses first along *y*, the particle is on the camera side of the focal plane. Both cases can be seen in the scans of the single particle included in figure 1.

## 3. Calibration

In order to accurately position particles imaged under the distortion of the cylindrical lens, we require a calibration relating reconstructed position with a ground truth obtained through scanning of a particle field using the HAPTV system. In this section we present two experimental scans at different magnifications to demonstrate the calibration process and evaluate the accuracy of our method. For both experimental setups shown in figure 3, we use identical illumination systems consisting of a HeNe laser (REO Inc; 12 mW) which passes through a spatial filter (Edmund Optics) to eliminate high frequency noise in the beam profile. The filtered beam is then collimated to a 50 mm diameter gaussian beam using a pair of lenses. Additionally, we also include a neutral density filter (Newport Inc) to modulate the intensity of the laser reaching our imaging system. We image the beam using separate imaging lenses and a long focal length cylindrical lens (Thorlabs; $f$ = 1000 mm) onto a CMOS camera (2048×1088 pixels Flare 2M360-CL) at a rate of 20 frames/s. The two magnification cases correspond to imaging resolutions of 1 μm/pixel using a 5X objective (EO M Plan Apo) and 10 μm/pixel, achieved with a Nikon imaging lens (Nikkor 105 mm f/2.8).

We record holograms of 4 μm silver coated hollow glass particles on a glass slide which is positioned at 30 μm increments using a micrometer stage for the high magnification experiment. Similarly, for the low magnification case, we capture 25 μm polystyrene particles across a total depth of 60 mm using a motorized translation stage. The use of the motorized translation stage provides a large range of translation at a uniform speed (177 μm/s), within which we can specify a constant displacement (1 mm) through sampling the recorded images in time.

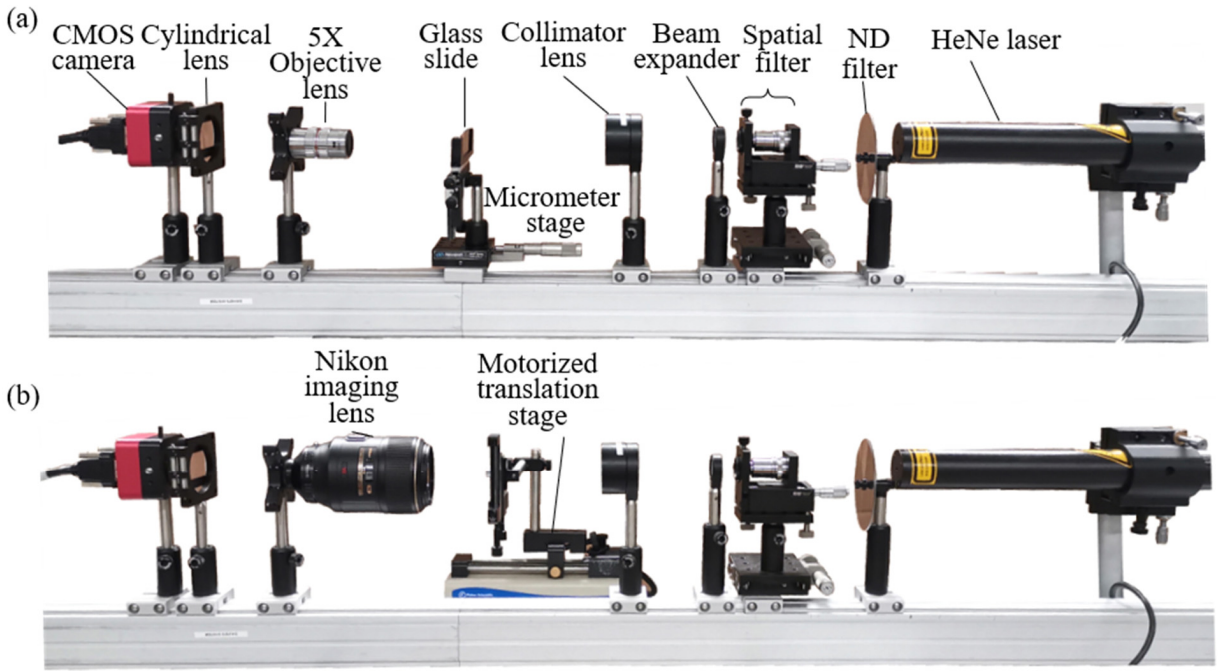

**Figure 3**. The experimental setup of HAPTV for the calibration of (a) high and (b) low magnification cases.

To determine a calibration curve, we use 50% of the recorded particles in our calibration experiment to fit a linear equation (see equations 1-2) relating the known and reconstructed positions for both magnifications (figure 4a and b). Both cases show a strong linear trend between the positions with slopes near to 1 which provides support for our depth definition of the particle using the focus position along the horizontal. As particles approach the focal plane ($z$=0) the paraxial approximation used in the diffraction kernel becomes invalid, preventing the recovery of the particle depth at such distances. Furthermore, the slopes of the calibration curves change with the image magnification, with the lower magnification being closer to unity compared to the other.

For the high magnification case:

$$z_{real} = 1.07 \times z_{detect} + 17.91 \ (\mu m) \quad (1)$$

For the low magnification case:

$$z_{real} = 0.99 \times z_{detect} - 388.10 \ (\mu m) \quad (2)$$

where, $z_{real}$ represents the ground truth positions and $z_{detect}$ represents the detected positions.

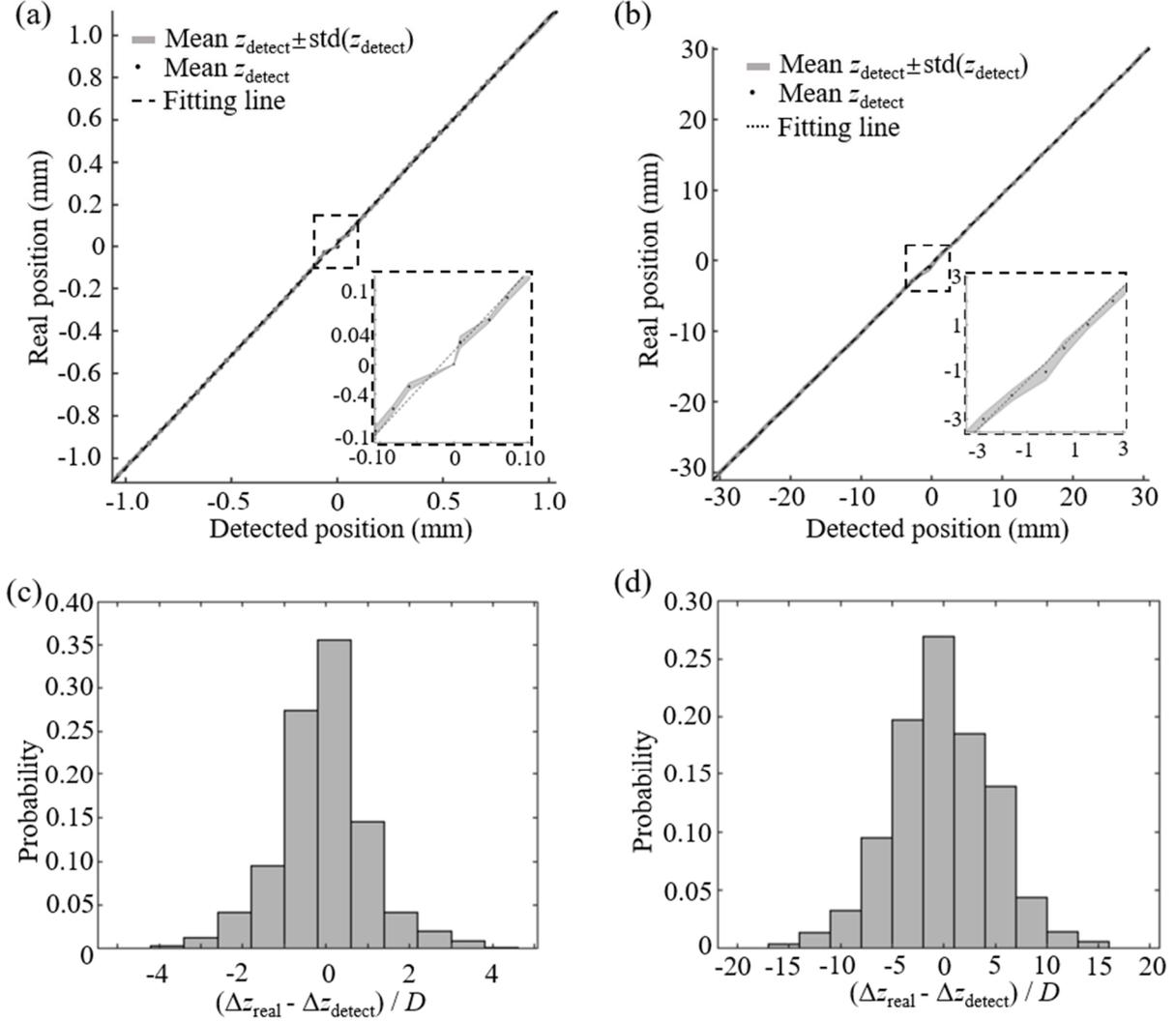

**Figure 4**. Calibration curve relating the real and detected positions for the HAPTV measurements at (a) high and (b) low magnification. The distribution of error in the calibration, given by the difference in displacements, for the (c) high and (d) low magnification cases.

To validate the accuracy of our calibration curve, we estimate an effective error using the remaining 50% of the recorded particles which were not used to generate the fit. We use the single step displacement in $z$ ($\Delta z_{detect}$) as our evaluation metric in order to eliminate a bias in the initial position for each individual particle from our measurement. A distribution of this displacement relative to the expected step size ($\Delta z_{real}$) for each particle, normalized by the particle diameter ($D$) is presented in figure 4(c) and 4(d) for each magnification. As illustrated, we are able to obtain an average error, represented by the standard deviation of the distribution, of ~1$D$ (4.2 μm) for the high magnification case and ~5$D$ (120.7 μm) for the low magnification case, respectively.

## 4. Validation through Measurements of 3D Jet Flow

Finally, we demonstrate our HAPTV technique by capturing a 3D jet flow and validate its accuracy through comparison to particle image velocimetry (PIV). As illustrated in figure 5, the experimental system consists of the flow chamber, the HAPTV setup, and a PIV imaging setup to capture the flow field in a plane within the sample volume of HAPTV. A 2 mm diameter nozzle placed at the bottom of an acrylic water tank (9×9×40 cm$^3$) is used to generate a laminar jet flow by injecting water from a syringe pump (Harvard Apparatus 22) at a constant flow rate of 0.25 ml/min from a 60 ml syringe (Monoject). We use neutrally buoyant 10 μm hollow glass spheres dispersed uniformly in the water as tracers for both experiments and include a 2 hour wait between the dispersal and measurement to ensure identical background flows in both cases. In order to ensure the jet has achieved a stable and repeatable flow profile, the syringe pump is started one minute prior to the start of the recording for both HAPTV and PIV. The flow is captured for a duration of 30 s with each technique. Furthermore, we perform three independent repetitions of the flow for each technique in order to obtain an accurate average velocity field for comparison.

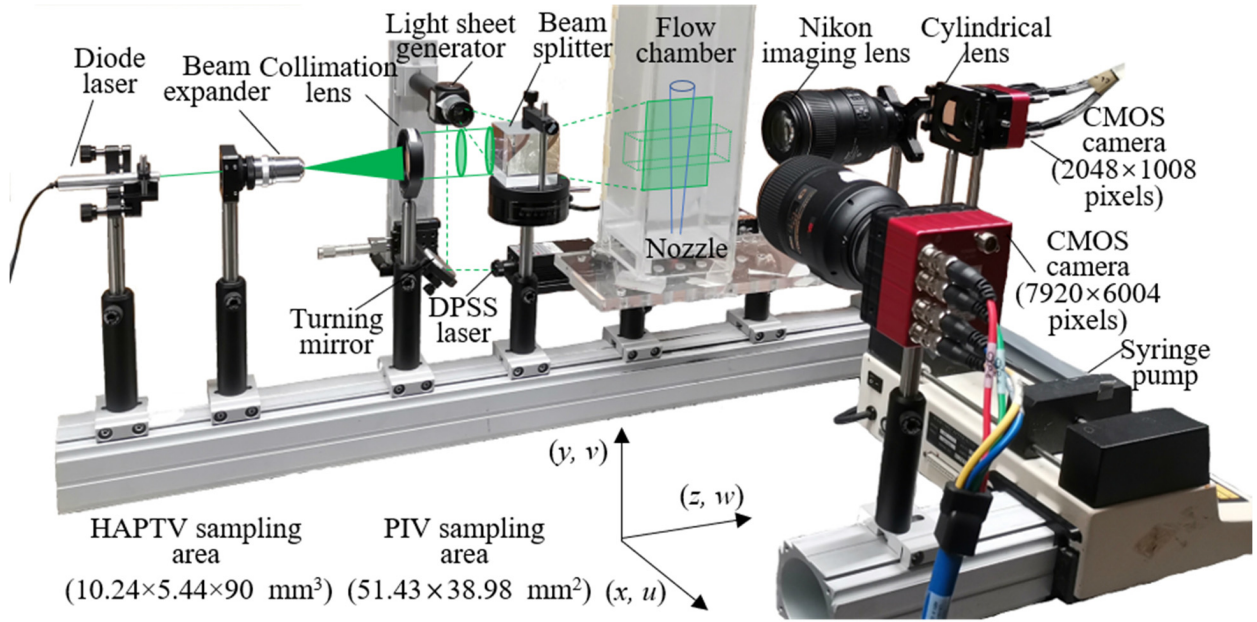

**Figure 5**. The experimental system for validating HAPTV through comparison to the PIV measurements of a jet flow.

The HAPTV system consists of a green diode laser (Thorlabs CPS532), a beam expander and a collimation lens, which together produce a 50 mm Gaussian beam to illuminate the sample volume. The motion of the tracers in the jet is imaged onto a CMOS camera (2048×1088 pixels Flare 2M360-CL) using a Nikon imaging lens (Nikkor 105 mm f/2.8) and a cylindrical lens ($f$ = 1000 mm) at a resolution of 5 μm/pixel recording at 320 frames/s with an exposure time of 61 μs. To demonstrate our ability to measure the entire flow field with the image plane located at an arbitrary position within the volume, we place the hologram focal plane with an offset of ~6.5 mm from the jet centerline. The PIV setup uses a green DPSS laser (Optoengine MGL-III-532-100), a turning mirror, and a light sheet generator to create a ~1 mm thick laser sheet which covers the $yz$ plane along the center of the jet. The PIV images are sampled by a separate CMOS camera (7920×6004 pixels, Flare 48MP) using a Nikon imaging lens (Nikkor 105 mm f/2.8) at a resolution of 6.5 μm/pixel recording at 30 frames/s with an exposure time of 1 ms. A beam splitter is placed in the optical path to position the laser sheet, ensuring that the Gaussian beam and laser sheet are coincident with each other and also ensuring that the two setups can capture the flow field at the same sampling region. In addition, we use a calibration target captured with both setups simultaneously as well as an image of the light sheet on the HAPTV camera to align the interrogation domains from both measurements. The image of the target on the two setups allows us to fix the vertical positions ($y$) within the volume to be identical while the image of the light sheet on the HAPTV sensor identifies the lateral ($x$) position on the hologram that corresponds to the PIV measurement location.

We use a multi-pass cross correlation approach using a 64×64 pixel interrogation window with 50% overlap to calculate the PIV vector fields (figure 6a) using PIVlab [40]. For HAPTV, the detected particles are first tracked over time to obtain an unstructured 3D velocity field. This field is then interpolated onto a 3D grid using inverse distance weighting and a grid resolution similar to PIV. This grid is the mean velocity field (figure 6b). We perform a separate ground truth calibration in water, since the resolution of the HAPTV measurement is increased for the current experiment relative to the prior low magnification calibration.

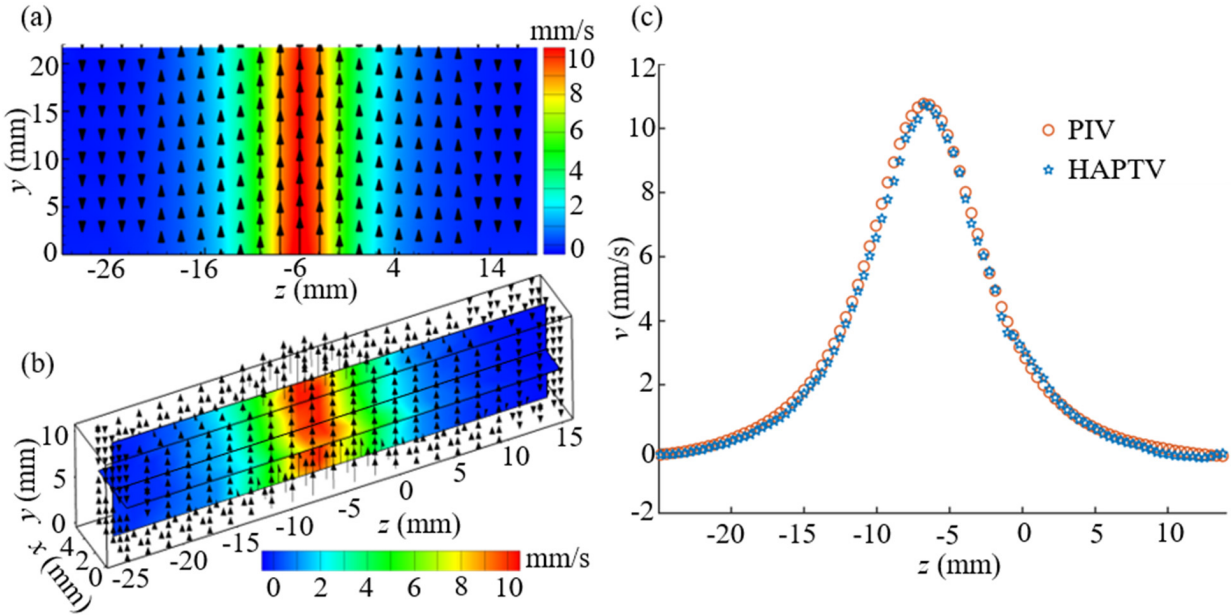

**Figure 6**. (a) Mean velocity field of vertical velocity ($v$) measured from the PIV and (b) HAPTV. (c) Comparison of mean velocity profiles from the PIV and HAPTV.

The flow profiles indicated in the PIV and HAPTV vector fields illustrate a typical jet flow profile with a peak velocity of 10 mm/s at the center that decays to zero velocity on either side. To offer a quantitative comparison between the two, we plot longitudinal ($z$) velocity profiles from both at $y = 10$ mm. For HAPTV, the profile is averaged over the thickness of the light sheet. The profiles are symmetric about the centerline even for HAPTV imaged away from the middle of the jet. The maximum difference between the two is ~5.5% of the centerline velocity and the full width at half maxima for the PIV and HAPTV measurements are 9.05 mm and 9.04 mm, respectively. The small difference between two measurements maybe caused by the misalignment of the jet flow and PIV light sheet.

## 5. Conclusion

In this study we introduce a novel flow measurement technique, astigmatic holographic PTV(HAPTV), by inserting a cylindrical lens in the setup of DIH-PTV which enables the image plane to be located in the middle of a sample volume while still differentiating the relative directions of particles from the image plane. Based on the holograms generated in this astigmatic system, we utilize a customized reconstruction algorithm with an intensity-based focus metric to distinguish tracers positioned on different sides of the focal plane. We perform a ground truth calibration of the measurement using a linear fit between the detected and true positions at two different magnifications. Our approach is calibrated under high (1 µm/pixel) and low (10 µm/pixel) magnifications with an error standard deviation of 4.2 µm (one particle size) and 120.7 µm (~5 times the particle size), respectively. As a demonstration, our method is implemented for the measurement of a 3D jet flow. The results show a good agreement when compared with the conventional PIV method. Overall, the present work demonstrates the capability of HAPTV to achieve improved depth of field compared to the conventional astigmatic PTV with the potential to perform flow measurement in field applications.

## Acknowledgements

This work is supported by the Fundamental Research Funds for the Central Universities (201822007). Zhou Zhou is supported by the Chinese Scholarship Council. We would also like to acknowledge the Qingdao National Laboratory for Marine Science and Technology for the support of experimental equipment. We would also like to thank University of Minnesota MnDrive program to provide some funding support for Santosh Kumar and Kevin Mallery.

## References

[1]    Wyngaard J C 1981 Cup, propeller, vane, and sonic anemometers in turbulence research *Annu. Rev. Fluid Mech.* **13** 399–423

[2]    Elderfield H and Schultz A 1996 Mid-ocean ridge hydrothermal fluxes and the chemical composition of the ocean *Annu. Rev. Earth Planet. Sci.* **24** 191–224

[3]    Kajitani L and Dabiri D 2005 A full three-dimensional characterization of defocusing digital particle image velocimetry *Meas. Sci. Technol.* **16** 790–804

[4]    Blanckaert K and Lemmin U 2006 Means of noise reduction in acoustic turbulence measurements *J. Hydraul. Res.* **44** 3–17

[5]    McLelland S J and Nicholas A P 2000 A new method for evaluating errors in high-frequency ADV measurements *Hydrol. Process.* **14** 351–66

[6]    Grant I 1997 Particle image velocimetry: A review *Proc. Inst. Mech. Eng. Part C J. Mech. Eng. Sci.* **211** 55–76

[7] Fu S, Biwole P H and Mathis C 2015 Particle tracking velocimetry for indoor airflow field: A review *Build. Environ.* **87** 34–44

[8] Doh D H, Cho G R and Kim Y H 2012 Development of a tomographic PTV *J. Mech. Sci. Technol.* **26** 3811–9

[9] Kim H, Westerweel J and Elsinga G E 2012 Comparison of Tomo-PIV and 3D-PTV for microfluidic flows *Meas. Sci. Technol.* **24** 24007

[10] Kitzhofer J and Brücker C 2010 Tomographic particle tracking velocimetry using telecentric imaging *Exp. Fluids* **49** 1307–24

[11] Belden J, Truscott T T, Axiak M C and Techet A H 2010 Three-dimensional synthetic aperture particle image velocimetry *Meas. Sci. Technol.* **21** 125403

[12] Deem E A, Zhang Y, Cattafesta L N, Fahringer T W and Thurow B S 2016 On the resolution of plenoptic PIV *Meas. Sci. Technol.* **27** 84003

[13] Chen H and Sick V 2017 Three-dimensional three-component air flow visualization in a steady-state engine flow bench using a plenoptic camera *SAE Int. J. Engines* **10** 625–35

[14] Troutman V A and Dabiri J O 2018 Single-camera three-dimensional tracking of natural particulate and zooplankton *Meas. Sci. Technol.* **29** 75401

[15] Barnkob R, Kähler C J and Rossi M 2015 General defocusing particle tracking *Lab Chip* **15** 3556–60

[16] Willert C E and Gharib M 1992 Three-dimensional particle imaging with a single camera *Exp. Fluids* **12** 353–8

[17] Verrier N, Coëtmellec S, Brunel M, Lebrun D and Janssen A J E M 2008 Digital in-line holography with an elliptical, astigmatic Gaussian beam: wide-angle reconstruction *JOSA A* **25** 1459–66

[18] Toloui M, Mallery K and Hong J 2017 Improvements on digital inline holographic PTV for 3D wall-bounded turbulent flow measurements *Meas. Sci. Technol.* **28** 44009

[19] Katz J and Sheng J 2009 Applications of Holography in Fluid Mechanics and Particle Dynamics *Annu. Rev. Fluid Mech.* **42** 531–55

[20] Yu X, Hong J, Liu C and Kim M K 2014 Review of digital holographic microscopy for three-dimensional profiling and tracking *Opt. Eng.* **53** 112306

[21] Mantovanelli A and Ridd P V. 2006 Devices to measure settling velocities of cohesive sediment aggregates: A review of the in situ technology *J. Sea Res.* **56** 199–226

[22] Graham G W and Nimmo Smith W A M 2010 The application of holography to the analysis of size and settling velocity of suspended cohesive sediments *Limnol. Oceanogr. Methods* **8** 1–15

[23] Kumar S S, Li C, Christen C E, Hogan Jr C J, Fredericks S A and Hong J 2019 Automated droplet size distribution measurements using digital inline holography *J. Aerosol Sci.* **137** 105442

[24] Shao S, Li C and Hong J 2019 A hybrid image processing method for measuring 3D bubble distribution using digital inline holography *Chem. Eng. Sci.* **207** 929–41

[25] Pereira F and Gharib M 2002 Defocusing digital particle image velocimetry and the three-dimensional characterization of two-phase flows *Meas. Sci. Technol.* **13** 683–94

[26] Yoon S Y and Kim K C 2006 3D particle position and 3D velocity field measurement in a microvolume via the defocusing concept *Meas. Sci. Technol.* **17** 2897–905

[27] Kim K C 2012 Advances and applications on micro-defocusing digital particle image velocimetry (μ-DDPIV) techniques for microfluidics *J. Mech. Sci. Technol.* **26** 3769–84

[28] Cierpka C and Kähler C J 2012 Particle imaging techniques for volumetric three-component (3D3C) velocity measurements in microfluidics *J. Vis.* **15** 1–31

[29] Chen S, Angarita-Jaimes N, Angarita-Jaimes D, Pelc B, Greenaway A H, Towers C E, Lin D and Towers D P 2009 Wavefront sensing for three-component three-dimensional flow velocimetry in microfluidics *Exp. Fluids* **47** 849–63

[30] Cierpka C, Segura R, Hain R and Kähler C J 2010 A simple single camera 3C3D velocity measurement technique without errors due to depth of correlation and spatial averaging for microfluidics *Meas. Sci. Technol.* **21** 45401

[31] Hain R, Kähler C J and Radespiel R 2009 Principles of a Volumetric Velocity Measurement Technique Based on Optical Aberrations *Imaging Measurement Methods for Flow Analysis* ed W Nitsche and C Dobriloff (Berlin, Heidelberg: Springer Berlin Heidelberg) pp 1–10

[32] Hsu W-Y, Lee C-S, Chen P-J, Chen N-T, Chen F-Z, Yu Z-R, Kuo C-H and Hwang C-H 2009 Development of the fast astigmatic auto-focus microscope system *Meas. Sci. Technol.* **20** 45902

[33] Rossi M and Kähler C J 2014 Optimization of astigmatic particle tracking velocimeters *Exp. Fluids* **55** 1–13

[34] Cierpka C, Rossi M, Segura R, Mastrangelo F and Kähler C J 2012 A comparative analysis of the uncertainty of astigmatism-μPTV, stereo-μPIV, and μPIV *Exp. Fluids* **52** 605–15

[35] Hinsberg N P Van, Roisman I V and Tropea C 2008 Three-dimensional, three-component particle imaging using two optical aberrations and a single camera *14th Int Symp Appl. Laser Tech. to Fluid Mech.* 7–10

[36] Yamaguchi I and Zhang T 1997 Phase-shifting digital holography *Opt. Lett.* **22** 1268–70

[37] Lai S, King B and Neifeld M A 2000 Wave front reconstruction by means of phase-shifting digital in-line holography *Opt. Commun.* **173** 155–60

[38] Ling H and Katz J 2014 Separating twin images and locating the center of a microparticle in dense suspensions using correlations among reconstructed fields of two parallel holograms *Appl. Opt.* **53** G1-11

[39] Matsushima K, Schimmel H and Wyrowski F 2003 Fast calculation method for optical diffraction on tilted planes by use of the angular spectrum of plane waves *J. Opt. Soc. Am. A* **20** 1755–62

[40] Thielicke W 2018 PIVlab–particle image velocimetry (PIV) tool